\documentclass[12pt]{article}
\usepackage{graphicx}
\usepackage[figuresright]{rotating}

\newcommand{\AmS}{{\protect\the\textfont2
  A\kern-.1667em\lower.5ex\hbox{M}\kern-.125emS}}
\newcommand{\secret}[1]{{}}

\newcommand\TeVcc{\ensuremath{\mbox{TeV}/c^2}}
\newcommand\TeV{\ensuremath{\mbox{TeV}}}
\newcommand\ttbar{\ensuremath{t\bar{t}}}
\newcommand\PYTHIA{PYTHIA}
\newcommand\GEANT{GEANT}
\newcommand\fbinv{\mbox{fb$^{-1}$}}
\newcommand\href[2]{#2}

\begin{document}
\begin{titlepage}
\title{Studies of Drell-Yan dimuon events in the CMS experiment}
\author{Alexander Lanyov\footnote{E-mail: Alexander.Lanyov@cern.ch} 
and Sergei Shmatov\footnote{E-mail: Sergei.Shmatov@cern.ch} \\
on behalf of the CMS Collaboration \\
\it Laboratory of Particle Physics, Joint Institute for Nuclear
Research, \\ \it 141980 Dubna, Russia}
%\date{}
\maketitle
\begin{abstract}
The potential of the Compact Muon Solenoid (CMS) experiment to
measure Drell-Yan muon pairs is discussed.
 Muon pairs can be measured in CMS with high precision up to very
high invariant masses.
 The systematic errors are considered.
 The potential to carry out precise measurements of the
forward-backward asymmetry is discussed.
\end{abstract}

\vspace{1cm}
\begin{center}
\small Presented at Hadron Collider Physics Symposium, 20-26 May 2007 \\
La Biodola, Isola d'Elba, Italy
\end{center}

\end{titlepage}

The Standard Model (SM) has been tested by the experiments at LEP,
SLC and Tevatron with high accuracy. 
  Extending these test at the new energy scale available with LHC is 
one of the priority tasks for particle physics.
  The results presented here are based on~note \cite{Belo06a} (see also \cite{phtdr2}),
which extends the studies for the LHC SM workshop 
(see~\cite{Hayw99} and references therein), using 
large samples of simulated events and the
CMS full detector simulation and reconstruction.
  In the Standard Model, the production of lepton pairs in hadron-hadron
collisions, the Drell-Yan (DY) process, is
described by $s$-channel exchange of photons or $Z$ bosons.  
  For measuring the mass dependence of Drell-Yan cross section 
one should know the detection efficiency, acceptance, resolution and total luminosity. 
  Simulation of Drell-Yan events in proton-proton collisions at 14
TeV centre-of-mass energy is performed with
\PYTHIA{} 6.217 using the CTEQ5L parton distribution functions.
  To simulate the detector geometry, materials and particle
propagation inside the detector, the \GEANT4-based simulation of
the CMS detector is used.
  The trigger simulation is based on the on-line reconstruction algorithms
selecting single- and double-muon triggers.
The total efficiency of triggering including
reconstruction and trigger selection efficiency is 98\,\% at 1\,\TeV. 
The additional cuts on calorimeter and tracker isolation of muon
tracks are not applied at High Level Trigger.

\begin{figure}[!Hhbt]\centering
    \includegraphics[width=0.48\textwidth]{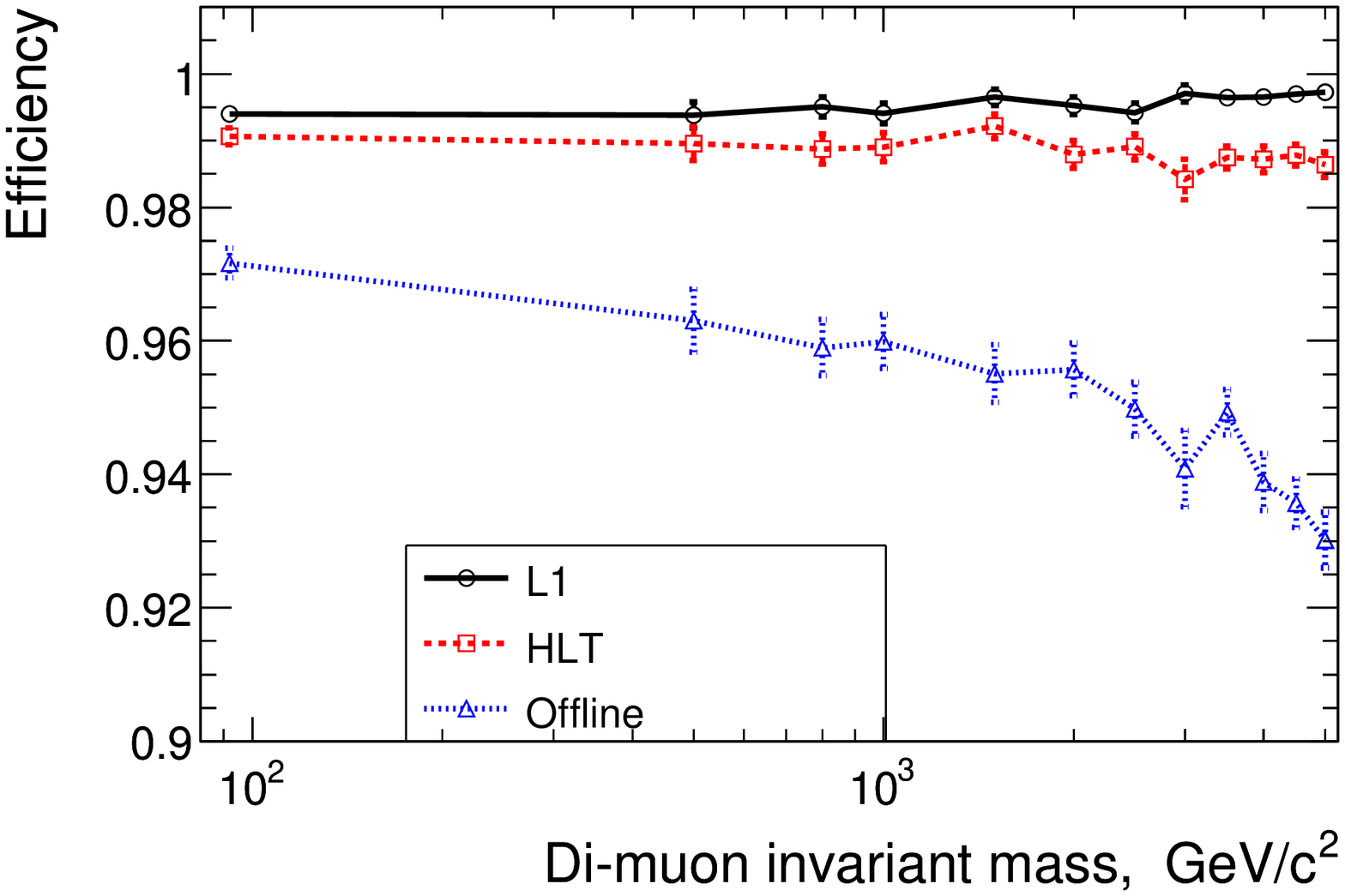}
    \includegraphics[width=0.48\textwidth]{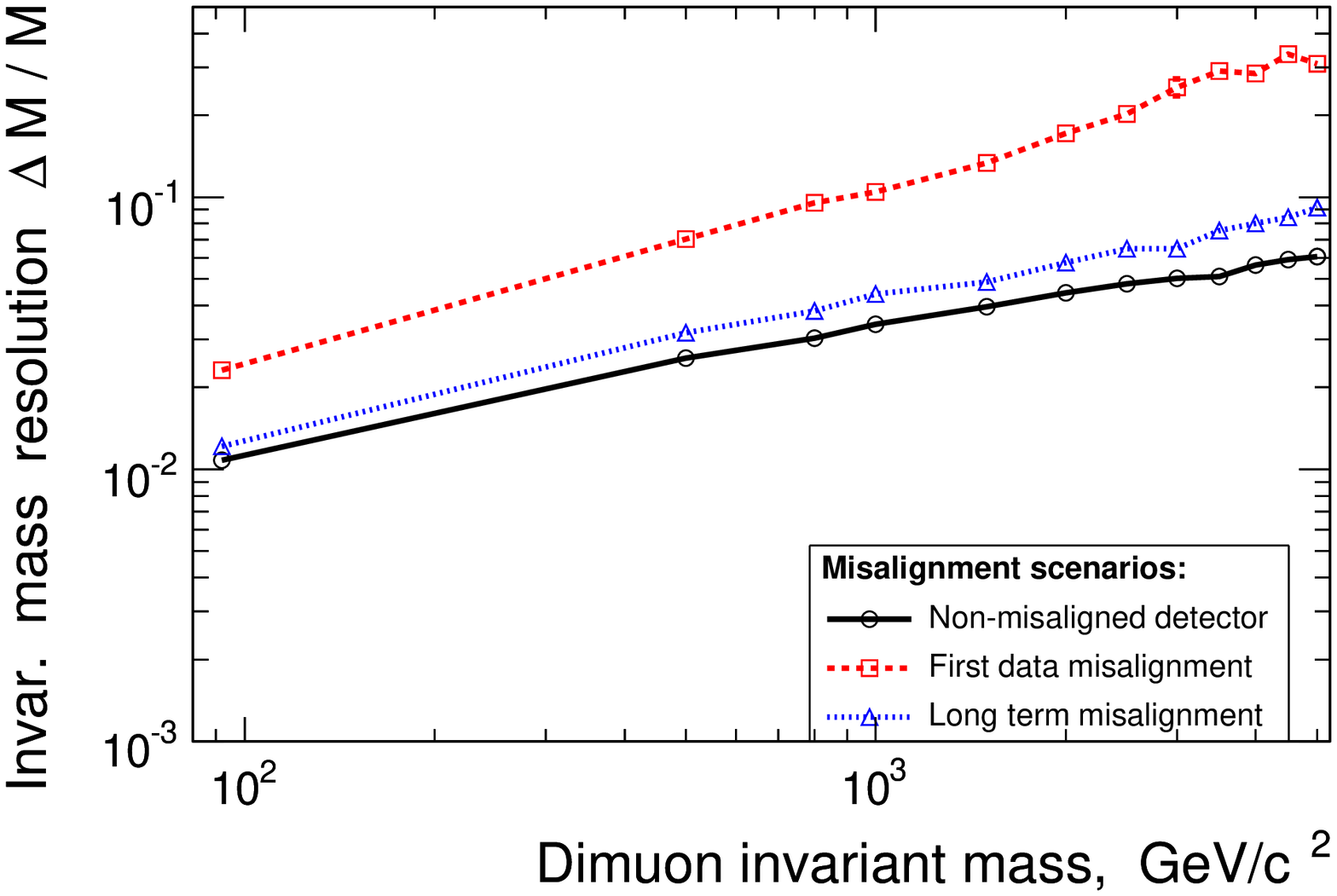}
    \caption{Left plot: dimuon reconstruction effi\-cien\-cy; 
             Right plot: invariant mass resolution; both as a function of the invariant mass cut.
             \label{fig:DY-reco}}
\end{figure}

The off-line muon reconstruction algorithm
is applied only to events which have passed trigger selection. At
the off-line level two muons inside the CMS acceptance
\mbox{$|\eta| \le$ 2.4} are required.  The overall efficiency of
the full reconstruction procedure taking into account trigger and
off-line reconstruction inefficiency is between 97\,\% and  93\,\% for
a mass range of 0.2 to 5 \TeVcc, as shown in the upper plot of 
Figure~\ref{fig:DY-reco}.  In the case of an ideal detector
the mass resolution smearing for fully-reconstructed events is
between 1.8\,\% and 6\,\% for the same mass range
(see the lower plot of Figure~\ref{fig:DY-reco}).  
 The effect of misalignment on the
mass resolution varies from 1.1\,\% up to 2.3\
depending on the level of misalignment
at the $Z$ and from 5\,\% up to 25\,\% 
for 3\,\TeVcc{} \cite{Belo06b}.

The 
backgrounds considered are vector boson pair
production $ZZ$, $WZ$, $WW$, \ttbar{} production etc. The
simulation and pre-selection of background events is done with the
same cuts as for the signal above. In the SM the expected
leading-order cross section of these events is negligible in
comparison with the Drell-Yan \cite{Belo06a}.
  The $\tau\bar{\tau}$ background (from $\tau$ decaying to $\mu$ and
neutrinos) is 0.8\,\% at the $Z$ pole and 0.7\,\% for masses above 1\,\TeVcc.
The background from Drell-Yan production of $q \bar q$ pairs (mostly
semi-leptonic $b$ or $c$ decays) is 0.3\,\% at the $Z$ pole without applying
any isolation cuts and below 0.1\,\% for masses above 1\,\TeVcc.  The other
background sources are negligible.  If the need arises they can be
further suppressed by acoplanarity and isolation cuts in the tracker.

  The statistical errors for 1, 10 and 100~\fbinv{} runs 
and the systematic uncertainty due to detector effects and uncertainties in the theory
are given in Table~\ref{tab:DY-crs_uncert}. 
  One can see that the systematic uncertainty due to smearing of
the reconstructed dimuon mass leads to modification of the
cross section as a function of dimuon mass and does not exceed 2.9\,\%
which is reached for a mass of 3\,\TeVcc.
  The misalignment does not affect the efficiency of dimuon
reconstruction for any masses~\cite{Belo06a,Belo06b}. 
  Taking into account that the trigger efficiency changes 
from 98.5\,\% to 97\,\% for masses from 0.2 to 5\,\TeVcc, 
very conservatively we may assign half of this change, i.e., 0.75\,\%,
as systematic uncertainty.

\begin{table}[h]
\caption{Relative errors of the Drell-Yan muon
         pairs cross section measurements in the fiducial volume.}
\label{tab:DY-crs_uncert}
\begin{center}
\vspace*{0.9ex}
\begin{tabular}{@{}lccccc}
\hline
$M_{\mu^+\mu^-}$,\hspace*{-3ex}~ & Detector\hspace*{-4ex}~  & \multicolumn{3}{c}{Statistical} & Theor.\hspace*{-4ex}~ \\
\TeVcc\hspace*{-3ex}~ & smearing\hspace*{-4ex}~ & 1 fb$^{-1}$\hspace*{-4ex}~ & 10 \fbinv\hspace*{-4ex}~ & 100 \fbinv\hspace*{-5ex}~ & Syst.\hspace*{-4ex}~ \\ \hline
$\ge$ 0.2         & 8 $\cdot$$10^{-4}$& 0.025       & 0.008      & 0.0026      & 0.058  \\
$\ge$ 0.5         & 0.0014            & 0.11        & 0.035      & 0.011       & 0.037  \\
$\ge$ 1.0         & 0.0049            & 0.37        & 0.11       & 0.037       & 0.063  \\
$\ge$ 2.0         & 0.017             &             & 0.56       & 0.18        & 0.097  \\
$\ge$ 3.0         & 0.029             &             &            & 0.64        & 0.134  \\
\hline
\end{tabular}
\end{center}
\end{table}

An important ingredient in the cross section measurement is the
precise determination of the luminosity. A promising possibility
is to go directly to the parton luminosity~\cite{Ditt97} by using
the $W^{\pm}$ ($Z$) production of single (pair) leptons. New
estimates show that in this way the systematic error on
$\sigma_{DY}^{high \ Q^2}$ {relative} to $\sigma_Z$ can be
reduced to $\approx 5-12$\,\%~\cite{Bour06}.

  On the theory side we consider several sources of systematic
uncertainties. The possible contributions from higher-order terms
in the dimuon production cross section are taken into account by
using a $K$ factor of $1.30\pm0.05$ as calculated with the
program~\cite{Hamb90}. It is expected that the total value of
additional NNLO contributions does not exceed 8\,\%.
  The EW corrections change
the cross section by 10-20\,\% \cite{Hayw99,Baur02}. The calculation
\cite{Zyku05} of the weak radiative corrections to the Drell-Yan
processes due to additional heavy bosons contributions shows that
these corrections are about 2.9\,\% to 9.7\,\% for mass region between
0.2\,\TeVcc{} and 5\,\TeVcc.
  The phenomenological origin of PDF gives an additional systematic error 
due to the PDF-dependence of the acceptance efficiency.
  The changes in the acceptance efficiency estimated by using the PDF
sets CTEQ5L, CTEQ6L and MRST2001E are up to 0.5\,\%.
  The ambiguity in the acceptance efficiency due to internal PDF
uncertainties is larger, but less than 1.4\,\% for any mass region.
  The experimental measurements of Drell-Yan cross section allow to fix
these theoretical uncertainties.

\begin{figure}[hbt]
\begin{center}
\hspace*{-3mm}\includegraphics[width=0.5\textwidth]{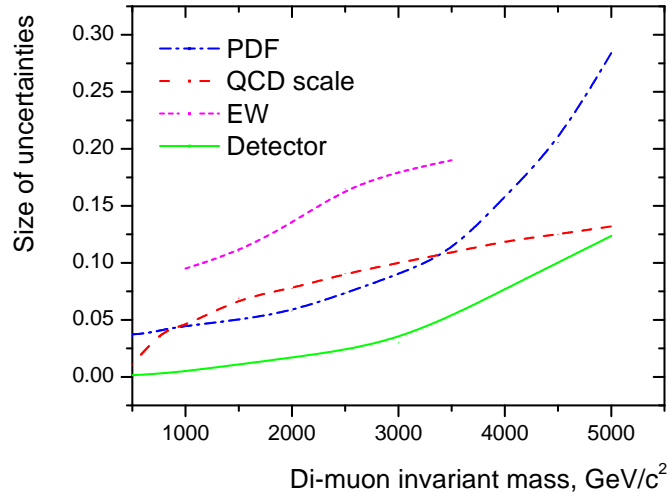}
\caption{Size of the EW corrections and the cross section uncertainties
         from PDFs, hard process scale and detector understanding as a
         function of the dimuon invariant mass cut.}
\label{fig:DY-uncert_sum}\end{center}
\end{figure}

The summary of the estimated systematic uncertainties as a function of
the dilepton mass is given in Figure~\ref{fig:DY-uncert_sum}.
Current uncertainties
from theory are larger than the experimental uncertainties.  The
statistical errors will dominate for invariant masses larger than 
2\,\TeVcc{} even for 100\,\fbinv.

The parton cross section in the lepton-pair centre-of-mass system has the form:
$$
 {1\over\sigma} {d\sigma \over d(\cos\theta^*)} =
 {3\over 2(3+b)} (1+b\cos^2\theta^*) + A_{FB}\cos\theta^*
$$
where $\cos\theta^*$ is angle between the outgoing negative lepton
and quark in the dilepton rest frame.
  To measure the forward-backward asymmetry $A_{FB}$ we need the original
quark and anti-quark directions of the initiating partons, but
these are not known in the case of $pp$ experiments, where the
initial state is symmetric. In Ref.~\cite{Rosn86,Ditt96} it is
shown that it is possible to approximate the quark direction with
the boost direction of the dimuon system with respect to the beam
axis. This is due to the fact that the valence quarks have on
average larger momentum than the sea anti-quarks, and therefore
the dimuon boost direction approximates the quark direction. The
most unambiguous tagging occurs for large dimuon rapidity.

  Without correction for mistags and acceptance 
the apparent $A_{FB}$ value will be twice smaller than the original
asymmetry ($\approx0.6$ for Drell-Yan events).
  However, using multi-dimensional fits~\cite{Cous05}
or reweighting techniques depending on the mistag and acceptance
which are under development, we can measure the original asymmetry.

The accuracy of asymmetry measurements depends on:
\begin{itemize}
  \item statistical uncertainty which grows with rising
the mass cut value, since the number of events for a given integrated luminosity 
$\int L\,dt=100\mbox{\,fb}^{-1}$ decreases with mass.
  \item systematic uncertainty from the variation of the mistag probabilities
for various PDF sets, typically below 10\,\%.
\end{itemize}
We expect the systematic uncertainty to dominate the statistical one for
integrated luminosity of $\int{L}\,dt=100\mbox{\,fb}^{-1}$
and dimuon masses around 500~GeV/$c^2$,
while the statistical one to be more important for dimuon mass cuts
above 1000~GeV/$c^2$.

\def\baselinestretch{0.6}

\end{document}